\newif\ifjournal\journalfalse

\ifjournal
\documentclass[12pt,preprint]{aastex}
\else
\documentclass[iop]{emulateapj}
\submitted{Submitted 2010 March 4; accepted 2010 May 24; published 2010 June 4}
\journalinfo{The Astrophysical Journal Letters, 716:L214 - L218, 2010 June 20}
\fi

\usepackage{graphicx}
\usepackage{bm}
\renewcommand{\vec}[1]{\boldsymbol{#1}}

\renewcommand{\div}{\nabla \cdot}
\def\pop{Phys. Plasmas}
\def\jcp{J. Comput. Phys.}

\shorttitle{RESISTIVE RELATIVISTIC MAGNETOHYDRODYNAMIC RECONNECTION}
\shortauthors{ZENITANI ET AL.}

\begin{document}

\title{Resistive Magnetohydrodynamic Simulations of Relativistic Magnetic Reconnection}

\author{Seiji Zenitani, Michael Hesse, and Alex Klimas}
\affil{
NASA Goddard Space Flight Center, Greenbelt, MD 20771, USA;
Seiji.Zenitani-1@nasa.gov
}

\begin{abstract}
Resistive relativistic magnetohydrodynamic (RRMHD) simulations
are applied to investigate
the system evolution of relativistic magnetic reconnection.
A time-split Harten--Lan--van Leer method is employed.
Under a localized resistivity,
the system exhibits a fast reconnection jet
with an Alfv\'{e}nic Lorentz factor
inside a narrow Petschek-type exhaust.
Various shock structures are resolved in and around the plasmoid
such as the post-plasmoid vertical shocks and
the ``diamond-chain'' structure
due to multiple shock reflections. 
Under a uniform resistivity,
Sweet--Parker-type reconnection slowly evolves.
Under a current-dependent resistivity,
plasmoids are repeatedly formed in an elongated current sheet.
It is concluded that the resistivity model is of critical importance
for RRMHD modeling of relativistic magnetic reconnection.
\end{abstract}

\section{Introduction}

Magnetic reconnection \citep{sweet,parker,petschek} is the driver of explosive events
in space, astrophysical, and laboratory plasmas.
The reconnection process attracts growing attentions
to explain flaring events \citep{lyut06,gia09} and
the magnetic annihilation \citep{coro90,lyu01}
in relativistic plasma environments. 
Basic properties of
relativistic magnetic reconnection have been discussed by
relativistic magnetohydrodynamic (RMHD) theories \citep{bf94b,lyu05,ten10}.
In particular, \citet{lyu05}'s careful work brought significant insights.

In the last decade,
modern simulation works have revealed
many features of relativistic magnetic reconnection.
From the viewpoint of kinetic physics,
it is widely recognized that
reconnection is an efficient particle accelerator
\citep{zeni01,zeni07,claus04}.
It was further found that the system is crucially influenced by
the guide-field magnetic topology \citep{zeni05} and
the radiative cooling effects \citep{claus09}. 
In a fluid scale, the reconnection system has been explored by
a resistive RMHD (RRMHD) \citep{naoyuki06} and
relativistic two-fluid \citep{zeni09a,zeni09b} models.
\citet{zeni09a} demonstrated and extensively analyzed
a quasi-steady Petschek-type reconnection.

There is a strong demand for
further development of RRMHD reconnection work.
Unlike the kinetic and two-fluid models,
the RRMHD model is free from kinetic scales
such as the skin depth and the gyro radius,
and so
it is highly desirable to study stellar-scale problems.
In addition, plasmas are considered to be collisional
in optically thick, radiation-dominated environments \citep{uz06}.
Since such plasmas can be approximated by a single RMHD fluid,
the RRMHD code with non-FLD\footnote{FLD is flux-limited diffusion} radiative transfer
will be necessary for future modeling of
radiative relativistic reconnection.

However, no other RRMHD work has come out since \citet{naoyuki06},
because the RRMHD equations turned out to be numerically unstable.
Attempts to better solve the RRMHD system have been undertaken.
\citet{kom07} pointed out that the non-ideal electric field behaves
as stiff relaxation to the ideal RMHD condition.
He split the equations into two parts:
the stiff terms were analytically solved as an exponential decay, and
the rest part was solved by a two-step Harten--Lan--van Leer (HLL) method.
\citet{pal09} employed a hybrid scheme of
an implicit Runge-Kutta method for the electric field and
an explicit one for the other variables.
\citet{dumbser09} developed
a higher-order scheme in unstructured volumes.

In this Letter, we present RRMHD simulations of
relativistic magnetic reconnection,
based on recent advances in numerical schemes \citep{kom07}.

\section{Numerical Setup}

We employ the following RRMHD equations in Lorentz--Heaviside notations with $c=1$.
\begin{eqnarray}
\label{eq:cont}
\partial_t (\gamma \rho) + \div (\rho \vec{u}) = 0 & \\
\label{eq:mom}
\partial_t (\vec{m} + \vec{E}\times\vec{B} )
+ \div \Big( ( p + \frac{B^2+E^2}{2} ) \vec{I} \nonumber \\
+ w \vec{u}\vec{u}
- \vec{B}\vec{B} - \vec{E}\vec{E} \Big)
&=& 0 \\
\label{eq:ene}
\partial_t ( \mathcal{E} + \frac{ {B}^2+{E}^2 }{2} )
+ \div ( \vec{m} + \vec{E}\times\vec{B} ) &=& 0
\end{eqnarray}
\begin{eqnarray}
\label{eq:B}
\partial_t\vec{B} + \nabla \times \vec{E} &=& 0 \\
\label{eq:E}
\partial_t\vec{E} - \nabla \times \vec{B} &=& -\vec{j} \\
\partial_t{\rho_c} + \div \vec{j} &=& 0 \\
\label{eq:ohm}
\gamma \Big( \vec{E} + \vec{v}\times\vec{B} - (\vec{E}\cdot\vec{v}) \vec{v} \Big)
&=& \eta ( \vec{j} - \rho_c \vec{v} )
\end{eqnarray}
In the above equations,
$\rho$ is the proper mass density,
$\vec{u}=\gamma\vec{v}$ is the spatial part of the 4-velocity,
$\vec{m}=\gamma w \vec{u}$ is the momentum,
$w$ is the enthalpy,
$\mathcal{E} = \gamma^2 w - p$ is the energy density,
$p$ is the proper pressure,
$\rho_c$ is the charge density, and
$\eta$ is the resistivity.
We employ a $\Gamma$-law equation of state
with an index of $\Gamma=4/3$
and so the enthalpy $w$ is given by $w = \rho+4p$. 
In addition, we use
a frame-independent magnetization parameter
$\sigma_{\varepsilon} = b^2/w = (B^2-E^2)/w$
and the conductivity $S=\eta^{-1}$
in our discussion.
$S$ also stands for the magnetic Reynolds number
based on the current sheet thickness $L=1$
and the typical speed of $c=1$, i.e., $S=Lc/\eta$.

\begin{deluxetable}{lccccccc}
\tabletypesize{\scriptsize}
\tablecaption{\label{table} List of Simulation Runs}
\tablewidth{0pt}
\tablehead{
\colhead{Run} &
\colhead{Domain} &
\colhead{Grids} &
\colhead{$\frac{\Delta t}{\Delta x}$} &
\colhead{$\rho_{in}$} &
\colhead{$\sigma_{\varepsilon,in}$} &
\colhead{$c_{A,in}$} &
\colhead{$\eta$}
}
\startdata
1 & $120 \times 120$ & $3600^2$ & 0.15 & 0.1 & 4.0 & 0.894 & Eq. \ref{eq:eta} \\
1L & $180 \times 180$ & $5400^2$ & 0.15 & 0.1 & 4.0 & 0.894 & Eq. \ref{eq:eta} \\
1b & $120 \times 120$ & $3600^2$ & 0.15 & 0.1 & 4.0 & 0.894 & Eq. \ref{eq:uni} \\
1c & $120 \times 120$ & $3600^2$ & 0.15 & 0.1 & 4.0 & 0.894 & Eq. \ref{eq:anom} \\
2 & $120 \times 120$ & $2400^2$ & 0.2 & 0.3 & 1.33 & 0.813 & Eq. \ref{eq:eta} \\
3 & $120 \times 120$ & $2400^2$ & 0.2 & 1.0 & 0.4 & 0.535 & Eq. \ref{eq:eta}
\enddata
\tablecomments{
Columns: (2) the domain size; (3) grid cells;
(4) the timestep ${\Delta t}/{\Delta x}$;
(5) the background density $\rho_{in}$;
(6) the relevant magnetization $\sigma_{\varepsilon,in}$;
(7) Alfv\'{e}n speed $c_{A,in}$; and
(8) the resistivity model.
}
\end{deluxetable}

We study two-dimensional system evolutions in the $x$--$z$ plane.
We employ a Harris-like model as an initial configuration:
$\vec{B} = B_0 \tanh(z)~\vec{\hat{x}}$,
$\vec{j} = {B_0} \cosh^{-2}(z)~\vec{\hat{y}}$, 
$\rho = \rho_0 \cosh^{-2}(z) + \rho_{in}$, $p=\rho$, $\rho_0=B_0^2/2=1$,
$\vec{u}=0$, and $\vec{E} = \eta \vec{j}$. 
The background density ($\rho_{in}$),
the resulting magnetization ($\sigma_{\varepsilon,in}$), and
the Alfv\'{e}n speed ($c_{A,in}$) are shown in Table \ref{table}.
In particular, $\sigma_{\varepsilon,in}$ is a measure of the system relativity,
because the reconnection outflow is expected to be
$\sim c_{A,in}=[\sigma_{\varepsilon,in}/(\sigma_{\varepsilon,in}+1)]^{1/2}$.
The reference parameters (run 1) are set similar to
those in the earlier works \citep{naoyuki06,zeni09a}.
We set the reconnection point at the origin $(x,z) = (0,0)$.
A small vector potential
$\delta A_y = 0.06 B_0 \exp[-(x^2+z^2)/4]$ is imposed
to trigger reconnection.
Because of the symmetry of the reconnection system,
we consider point-symmetric conditions at $x=0$:
a property $f$ satisfies $f(0,z) = f(0,-z)$ or $-f(0,-z)$. 
Neumann conditions of zero normal derivatives are employed
at inflow ($z=\pm 60$ [$\pm90$ in run 1L])
and outflow ($x=120~[180]$) boundaries. 
The normal magnetic fields ($B_z$) are adjusted to
ensure $\div \vec{B}=0$ at the inflow boundaries.

We solve the RRMHD equations by
a variant of \citet{kom07}'s time-split HLL method.
We solve a subset of the full equations,
because our specific configuration ensures
$B_y$, $E_x$, $E_z$, $u_y$, $\rho_c$, $j_x$, $j_z$, and $(\vec{E}\cdot\vec{v})=0$.
Primitive variables are analytically recovered \citep{zeni09a}
and then interpolated by
a second order monotonized central (MC) limiter.
We use the hyperbolic divergence cleaning method
for the solenoidal condition \citep{dedner02}.

We study three different resistivity models ---
uniform, spatially localized, and current-dependent ones.
\begin{eqnarray}
\label{eq:uni}
\eta & = & \eta_0,
\\
\label{eq:eta}
\eta(x,z) &=& \eta_0 + (\eta_1-\eta_0) \cosh^{-2}(\sqrt{x^2+z^2}),
\\
\label{eq:anom}
\eta(j^2,\rho) &=&
\left\{
\begin{array}{ll}
\eta_0
& ~~~~~ ( j^2 < \rho^2I^2_{c})
\\
\eta_0 \sqrt{j^2} ( \rho I_{c} )^{-1}
& ~~~~~ ( j^2 \ge \rho^2 I^2_{c} )
.
\end{array}
\right.
\end{eqnarray}
Here, $j^2=j^{\mu}j_{\mu}=\vec{j}^2-\rho_c^2$ is
a frame-independent variable and $I_c$ is a threshold.
The baseline value is $S_0 =\eta_{0}^{-1}=80$
in all three models.
We mainly explore the second model,
whose localized value is $S_1 =\eta_{1}^{-1}=10$.
The third model depends on the space-like current intensity,
and the anomalous resistivity starts to work
when the current is 5 times stronger than
the Harris current, $I_c = 5 B_0 / \rho_0$. 
These resistivity models are handled
by the analytic part of the numerical scheme. 
In the current-dependent case,
the partial Amp\`{e}re's law $\partial_t E_y = - j_y$
in the split part (see \citet{kom07}, section 3)
has the following solution:
\begin{eqnarray}
\label{eq:anom2}
j_y &=&
\left\{
\begin{array}{ll}
j_{y0} \exp( -S_0 \gamma t)
& ~~~ ( |j_y| < \rho I_{c})
\\
j_{y0} -
\frac{1}{2}({\rm sgn}j_{y0})
\rho I_c S_0 \gamma t
& ~~~ ( |j_y| \ge \rho I_{c})
,
\end{array}
\right.
\end{eqnarray}
where $j_{y0}$ is the initial state of $j_y$.

\section{Results}

\begin{figure*}[htbp]
\begin{center}
\ifjournal
\includegraphics[width={\columnwidth},clip]{f1.eps}
\else
\includegraphics[width={2\columnwidth},clip]{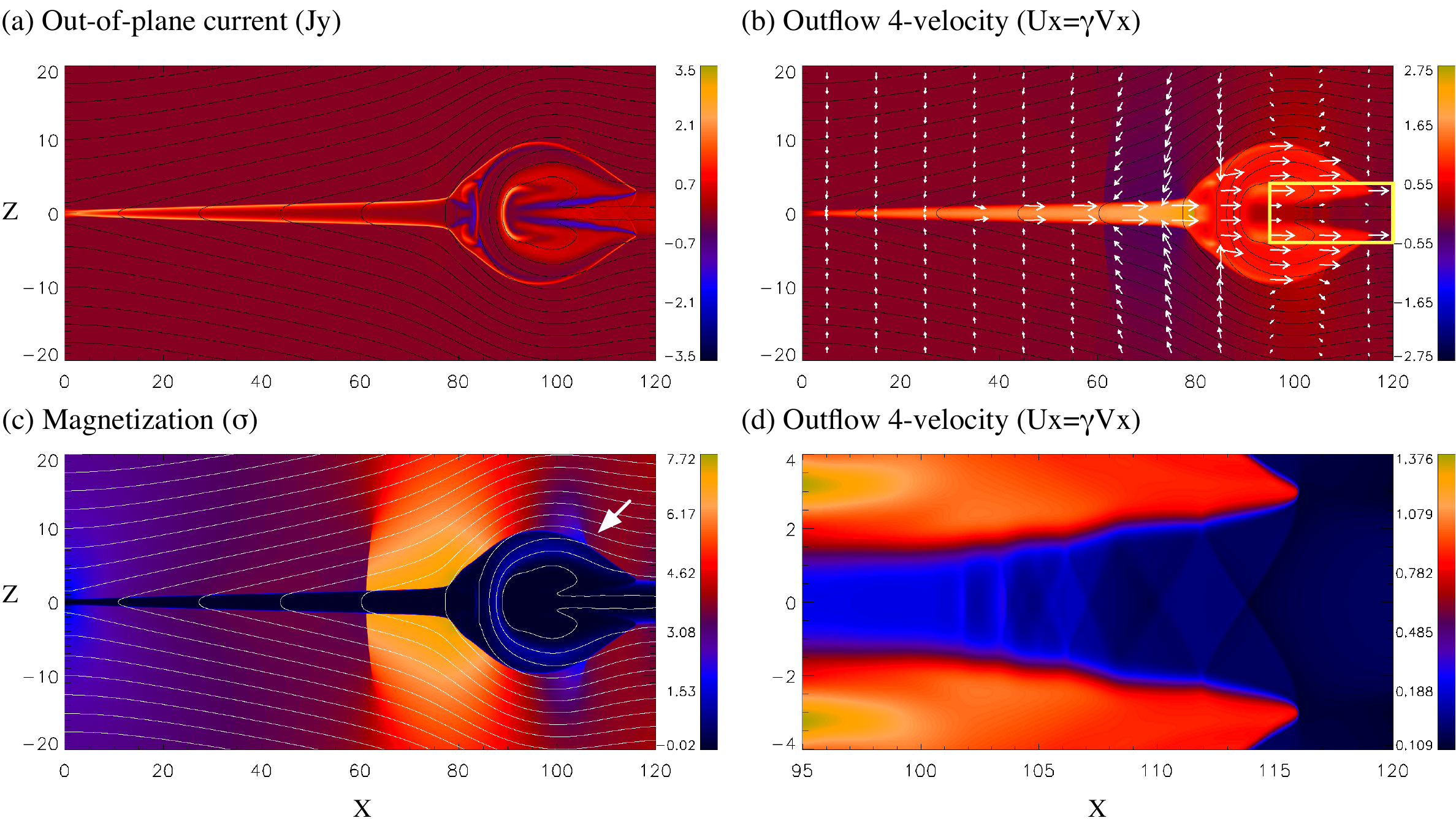}
\fi
\caption{
Snapshots of run 1 at $t=195$ in the $x$--$z$ plane.
Contour lines show the magnetic fields.
(a) The out-of-plane current $j_y/j_0$.
(b) The 4-velocity $u_x$ and the velocity vectors $\vec{v}$.
When $|v|=c$, the length of the arrow is 6. 
(c) The magnetization $\sigma_{\varepsilon}=(B^2-E^2)/w$.
(d) The 4-velocity $u_x$ in $95\le x\le 120$ and $-4 \le z \le 4$
(the small box in panel (b)).
\label{fig:snap}}
\end{center}
\end{figure*}

We study run 1
with the localized resistivity (Equation \ref{eq:eta}).
After an initial adjustment stage of $t \lesssim 10$,
the reconnection process starts around the origin, and then
it transfers the upstream magnetic fields faster and faster.
The electric field $E_y$ at the reconnection point (hereafter $E_y^{*}$),
which stands for the flux transfer speed of reconnection,
exceeds $E_y^* \sim 0.1 B_0$ at $t \sim 50$.
The normalized reconnection rate
$\mathcal{R} = E_{y}^{*}/(c_{A,in'}B_{in'}) \approx v_{in'}/v_{out}$
(using $v_{out} \approx c_{A,in'}$ and $E_{y,in'} \approx E_{y}^{*}$)
reaches its typical value of $\mathcal{R} \sim 0.145$ at $t \sim 100$.
Here, the upstream properties with the subscript $in'$ are evaluated at $(x,z)=(0,20)$. 
Around this time, we recognize key signatures that are discussed later in this section,
and main structures start to move at the nearly constant speeds \citep{ugai95}.
Therefore, we think that the reconnection is
well developed after $t \gtrsim 100$.

Figure \ref{fig:snap} shows the snapshots
at $t=195$.  Panels (a),(b), and (c) show
the out-of-plane current $j_y$,
the 4-velocity $u_x$, and
the magnetization $\sigma_{\varepsilon}$, respectively.
A fast reconnection jet travels outward
inside a narrow exhaust
between a pair of slow shocks \citep{petschek}.
The jet collides with a big magnetic island (the so-called plasmoid)
in front of the current sheet further downstream.
The typical outflow 4-velocity is $u_x\sim 1.8$ ($\gamma \sim 2$), and
its maximum value is $u_x = 2.75$ ($\gamma = 2.93$)
near the neck of the plasmoid ($x \sim 80$).
The proper density inside the outflow exhaust is
$3$ times bigger than the that in upstream region. 
The enthalpy flux ($4 \gamma p u_x$)
carries 90\% of outgoing energy flow at $x = 60$
due to the high pressure. 

Since the plasmoid suddenly compresses the surrounding plasmas,
the increased pressure pushes plasmas
in the field-aligned directions.
In the post-plasmoid regions ($60<x<80$),
the plasma inflow is pushed in the $-x$-direction
($u_x<0$; blue regions in Figure \ref{fig:snap}(b)).
In Figures \ref{fig:snap}(b) and (c),
a pair of vertical shocks can be seen at $x\sim 60$
between the left regions and
the post-plasmoid regions.
They are slow shocks and they move to the right
at a constant speed of ${\sim}0.41c$ in the well-developed stage. 
Note that the left side is the shock downstream and the right is the upstream.
In Figure \ref{fig:snap}(c), 
$\sigma_{\varepsilon}$ is enhanced
around the post-plasmoid regions.
This is due to the lower pressure,
caused by a field-aligned expansion of plasmas.
The maximum speed in the outflow exhaust is comparable with
Alfv\'{e}n speed of the enhanced $\sigma_{\varepsilon}$ ($\sim 7.7$)
around there.
Another pair of small shocks are found
outside the plasmoid at $x\sim 105$,
as indicated by the arrow in Figure \ref{fig:snap}(c).

The plasmoid is surrounded by strong positive currents
as can be seen in Figure \ref{fig:snap}(a).
They are slow shocks \citep{ugai95},
which are connected to the Petschek slow shocks.
Due to the plasma heating and the magnetic energy dissipation across those slow shocks,
the magnetization is weak inside the plasmoid
(Figure \ref{fig:snap}(c)).
One exception is the blue arc region,
where the reconnected magnetic fields ($B_z$) are piled up
at $x \sim 84$-$89$.
There is a tangential discontinuity
on the right vicinity of the piled-up region
at $x\sim 90$.
It looks like the small bright `C'-shaped structure in Figure \ref{fig:snap}(a).
There are a pair of reverse current structures around $x \sim 90$-$110$
(blue regions in Figure \ref{fig:snap}(a)).
The magnetic fields are so bent across these oblique structures
that their tangential components change the polarity.
They are intermediate shocks \citep{shuei01}. 

Figure \ref{fig:snap}(d) shows
the $u_x$-profile near the plasmoid edge.
The same domain is indicated by
the yellow box in Figure \ref{fig:snap}(b).
Plasmas are stationary around the central region ($z\sim 0$)
because the high-density Harris sheet plasmas are confined. 
Meanwhile, the twin plasmoid flows travel fast
in the upper and lower regions ($z\sim\pm 3$).
They are separated by the intermediate shocks
from the central region.
The plasmoid edges propagate at ${\sim}0.75c$ and
the plasma speed is ${\sim}0.66c$ there.
Interestingly, we find the chain of diamond-shaped features
around $x\sim 102$-$115$, $z\sim 0$.
We call them the ``diamond-chain'' structure. 
Since the twin plasmoid edges move faster
than the fast mode or the sound speed
inside the central region (${\sim}0.53c$),
bow shocks propagate from there,
and then the shock fronts are reflected
by the intermediate shocks multiple times. 
The diamond-chain starts to develop
in earlier stages of simulation and
Figure \ref{fig:snap}(d) shows that
six or more reflections occurred. 
We confirmed that the structure is virtually unaffected by the boundary effect
by carrying out a larger run (run 1L). 
In the right side, the current sheet looks
60-100\% wider than the initial condition.
This is due to the current sheet diffusion
whose timescale is $L^2S_0 \sim S_0 = 80$.

\begin{figure}[htbp]
\begin{center}
\ifjournal
\includegraphics[width={\columnwidth},clip]{f2.eps}
\else
\includegraphics[width={\columnwidth},clip]{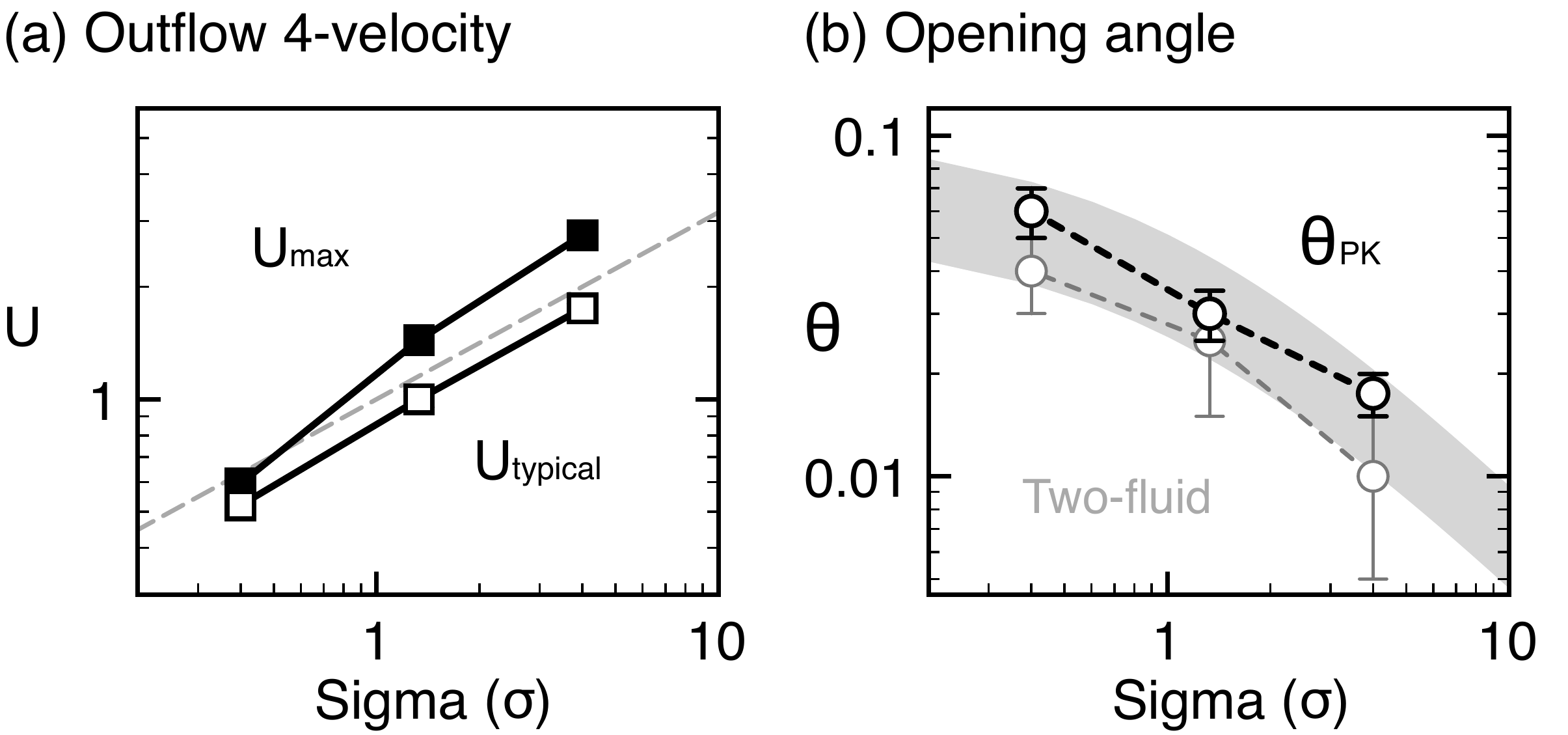}
\fi
\caption{
(a)
Maximum and typical 4-velocities $u_x$
as a function of the initial $\sigma_{\varepsilon,in}$.
The dashed line indicates an Alfv\'{e}nic value, $\sqrt{\sigma_{\varepsilon}}$.
(b)
Petschek opening angles (black line) and
the relevant results in two-fluid work \citep{zeni09a} (gray line).
The shadow shows $0.05 < (1+\sigma_{\varepsilon,in})~\theta < 0.1$.
\label{fig:sigma}}
\end{center}
\end{figure}

Next, we study the dependence on the upstream magnetization.
We compare runs 1-3 in their well-developed stages.
Typical reconnection rates are $\mathcal{R} = 0.145, 0.125$, and $0.102$,
respectively ---
reconnection evolves faster in higher-$\sigma_{\varepsilon}$ cases. 
Shown in Figure \ref{fig:sigma}(a) are
the 4-velocities in the outflow exhaust.
Since $u_x$ anomalously increases near the plasmoid neck,
we evaluate the typical 4-velocity
in the left side of the vertical shocks. 
The typical 4-velocities (white squares) are
in better agreement with
the Alfv\'{e}n value $\sqrt{\sigma_{\varepsilon}}$ (dashed line)
than the maximum (black squares).
We also find that
the upstream magnetization gradually changes during the system evolution.
For example, $\sigma_{\varepsilon}$ decreased from $4$ to $3.5$
around $x\sim 60$ in Figure \ref{fig:snap}(c).
Considering this, the agreement becomes even better.
This relation bridges
the well-known nonrelativistic result of $v_{out} \approx c_{A,in}$ and
the relativistic scaling in the high-$\sigma_{\varepsilon,in}$ limit,
$\gamma_{out} \sim \sigma_{\varepsilon,in}$ \citep{lyu05,zeni09a}.

Figure \ref{fig:sigma}(b) presents
the opening angle $\theta_{PK}$ of Petschek current layers. 
For reference, results of previous two-fluid works \citep{zeni09a}
are presented too. 
In both cases,
the angles become narrower
as $\sigma_{\varepsilon}$ increases. 
This intuitively fits the traditional explanation of the Petschek outflow.
In the jet frame,
the outflow exhaust expands by a local Alfv\'{e}n speed,
$\propto B'_z \sim B_z/\gamma_{out}$.
In the observer's frame, such transverse motion
in the relativistic jet
looks slower by another factor of $\gamma_{out}$.
As a result, given that all the other conditions are the same,
the Petschek angle is rescaled by
$\gamma^{-2}_{out} \approx (1+\sigma_{\varepsilon,in})^{-1}$.
Assuming that the slope angle of the upstream magnetic field is
$\theta_m \sim \mathcal{R}$, the Petschek angle would be
a sizable fraction of $\mathcal{R} / (1+\sigma_{\varepsilon,in})$.
For an order estimate, the shaded region in Figure \ref{fig:sigma}(b)
shows the range of $0.05 < (1+\sigma_{\varepsilon,in})~\theta < 0.1$. 
The angle analysis was further extended to $\sigma_{\varepsilon,in}=9$ and
compared with the high-$\sigma_{\varepsilon,in}$ theory
in the two-fluid work \citep{zeni09a}. 
To comprehensively discuss the outflow speed and the Petschek angle,
a generalized theory which covers
the full range from the nonrelativistic low-$\sigma_{\varepsilon,in}$ regime
to the relativistic high-$\sigma_{\varepsilon,in}$ limit
needs to be developed.

\begin{figure}[htbp]
\begin{center}
\ifjournal
\includegraphics[width={\columnwidth},clip]{f3.eps}
\else
\includegraphics[width={\columnwidth},clip]{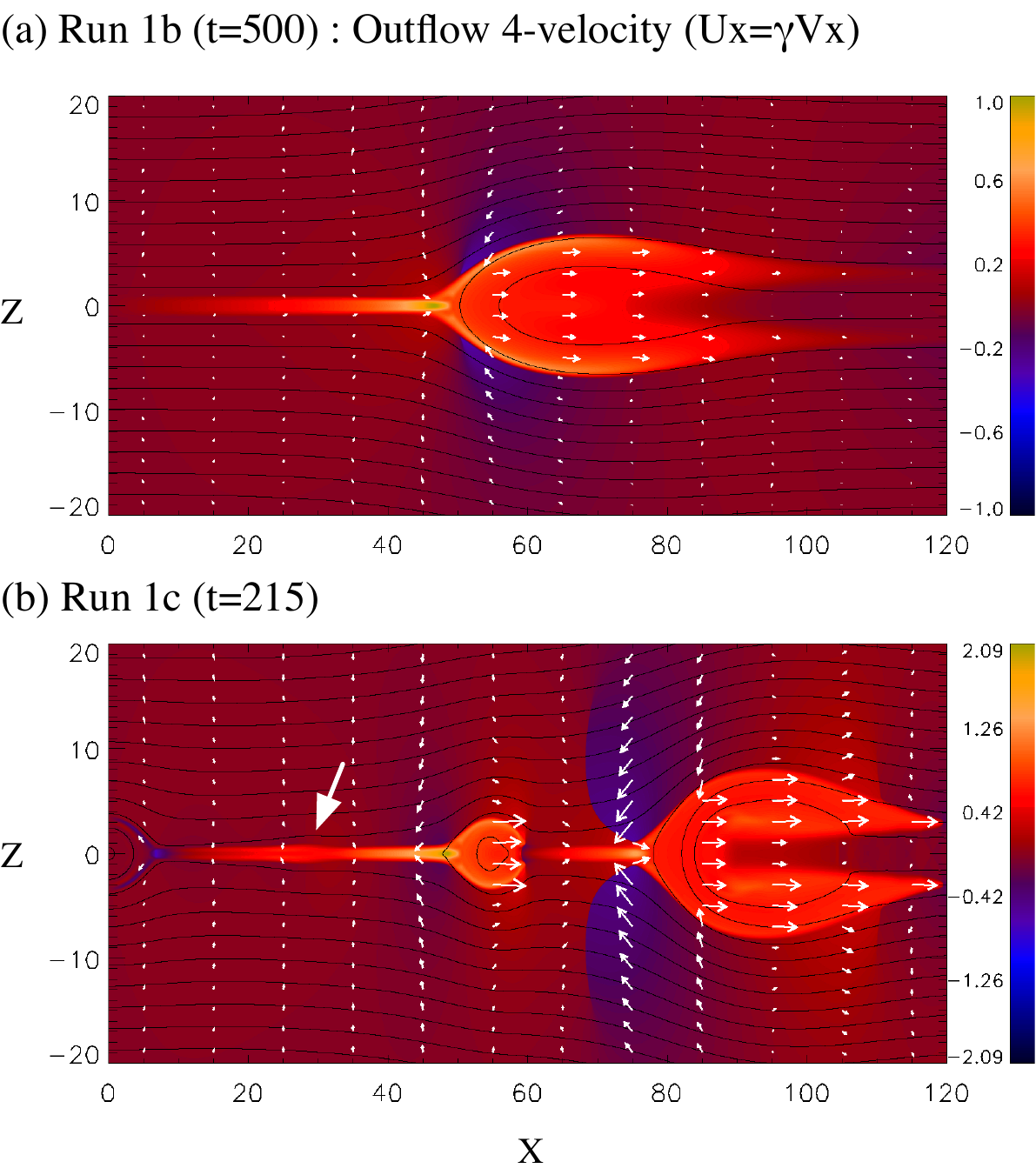}
\fi
\caption{
Snapshots for
(a) run 1b at $t=500$ and for
(b) run 1c at $t=215$:
the 4-velocity $u_x$ (color), 
the velocity $\vec{v}$ (arrows; the length is 6 when $|v|=c$), and
the magnetic fields (contour lines).
\label{fig:snap2}}
\end{center}
\end{figure}

We further study the effect of the resistivity model.
Run 1b employs the uniform resistivity (Equation \ref{eq:uni})
with the same initial conditions as run 1.
In this case, the system evolves very slowly.
The plasmoid structure is visible after $t\sim 400$ and
the outflow becomes relativistic $u_x\sim 1$ at $t= 475$-$500$.
Figure \ref{fig:snap2}(a) shows the snapshot at $t=500$.
The reconnected current sheet remains thin and
it is simply elongated in the $x$-direction:
the system exhibits a Sweet--Parker-type reconnection. 
The reconnection rate is slower by an order of magnitude,
$\mathcal{R}\sim 0.01$-$0.02$.
Obviously the localized resistivity
changed the system evolution. 
Run 1c employs the current-dependent resistivity (Equation \ref{eq:anom}). 
By this model,
we intend to limit the current in low-dense regions,
where the system does not have a lot of current carriers.
The run starts from the intermediate data of run 1 at $t=25$
to speed up the onset. 
In this case, a Sweet--Parker-type reconnection similarly grows,
but plasmoids repeatedly grow in a single elongated current sheet. 
Figure \ref{fig:snap2}(b) shows
a well-developed stage at $t=215$. 
One plasmoid stays near the origin
because of the symmetric boundary condition.
A new plasmoid starts to grow at $x\sim 29$
(indicated by the white arrow in Figure \ref{fig:snap2}(b)),
and the older one is ejected to the downstream, $x\sim 55$.
Such a repeated formation of plasmoids is
one of the ubiquitous features of reconnection. 
Their distances are comparable with
the typical wavelength of the tearing mode $\sim 20$-$30$. 
The diamond-chain develops around the edge of the biggest plasmoid,
although it is difficult to distinguish in Figure \ref{fig:snap2}(b).
The anomalous resistivity plays a role
both in the elongated current sheet and
at the slow shocks surrounding the post-plasmoid regions.
The conductivity changes $S \sim 80 \rightarrow 15$-$30$
due to the intense currents there.
Overall evolution is slower than that of run 1,
but faster than that of run 1b.
Using the electric field at the reconnection point around $x\sim 8$,
an equivalent reconnection rate is $\mathcal{R} \sim 0.05$-$0.06$.

\section{Discussion and Summary}

Our results corroborate
the first RRMHD work by \citet{naoyuki06} on many aspects
such as Alfv\'{e}nic outflow, the compression ratio, and
the narrower slow shock structure.
Judging from their inflow speed ($v_{in}/c_{A,in}\sim \mathcal{R}$),
they observed faster reconnection
probably due to the stronger localized resistivity of $S_1 \sim \mathcal{O}(1)$.
Our results also agree with the two-fluid work by \citet{zeni09a}
such as the typical outflow parameters of the reference runs and
the Petschek angle relations. 
Some visible differences are attributed to the two-fluid effects. 
The current layers look much sharper in this work,
because the fluid inertial effects tend to
smooth structures in the two-fluid model.
The plasmoid looks longer in the $x$-direction.
The outflow jet immediately hits the downstream plasmas
in the $x$-direction in the RRMHD model,
while plasma out-of-plane motion softens the reflection in the two-fluid model. 
Concluding this and those works,
(1) Alfv\'{e}nic outflow,
(2) increasing reconnection rate, and
(3) narrower opening angle,
can be regarded as common features of relativistic Petschek reconnection. 
Since a narrower outflow exhaust suppresses the energy throughput,
(3) does not immediately fit (2).
However, \citet{zeni09b} pointed out that
the relativistic enthalpy flux transports
huge energy per a cross section even in a narrower exhaust.

Thanks to the stable RRMHD code,
we successfully resolved shock structures in and around the plasmoid.
To our knowledge,
the post-plasmoid vertical shocks and
the diamond-chain structure are new discoveries.
The diamond chain develops
when the plasmoid edge speed (a sizable fraction of $c_{A,in}$)
exceeds the sound speed in the neutral Harris sheet.
We predict that it also develops
in nonrelativistic MHD simulations
once the above condition is met.
Since multiple shocks are confined in a narrow region,
the plasmoid edge can be a potential site of shock acceleration of particles.

From the numerical viewpoint,
relativistic reconnection in the high-$\sigma_{\varepsilon}$ regimes
is challenging for an explicit-type RRMHD code. 
Since the analytic part of the scheme contains
the time constant of $-S\gamma$ (e.g., Equation \ref{eq:anom2}a),
employing Alfv\'{e}nic outflow speed,
a code requires
$ \Delta t \lesssim S_0^{-1} (1+\sigma_{\varepsilon,in})^{-1/2}$. 
As shown in Section 3,
the outflow is faster near the plasmoid neck,
and so the restriction becomes even severer. 
Thus, high time (and appropriate spatial) resolutions are necessary
in high-$\sigma_{\varepsilon}$ and moderate-$S$ regimes of our interest. 
In addition, a plasma solution is not always stable
even when the total momentum and energy are well handled.
Let us extract the plasma momentum from Equation \ref{eq:mom},
$\partial_{t} \vec{m}
+ \div ( p\vec{I} + w \vec{u}\vec{u} )
= \vec{j} \times \vec{B}$.
Focusing on the source term which is usually problematic,
we see that the plasma stability condition,
${|\Delta \vec{m}|}/{w}
\sim {| \vec{j} \times \vec{B} | \Delta t }/{w} 
\lesssim \mathcal{O}(0.1),$
will be critical in higher-$\sigma_{\varepsilon}$ regimes,
unless Ohm's law strongly limits the current.
These issues need to be improved by an implicit approach \citep{pal09}. 

From the physics viewpoint,
those restrictions obviously come from an immature form of Ohm's law. 
Note that the RRMHD model contains
the displacement current $\partial_t\vec{E}$ in Amp\`{e}re's law (Equation \ref{eq:E}).
Since it operates in a short timescale of the plasma frequency,
the nonrelativistic MHD code usually drops it.
However, the RRMHD system needs it
for the energy and momentum conservation.
On the other hand, the RRMHD uses a simple Ohm's law
with the scalar resistivity (Equation \ref{eq:ohm}).
This is a time-stationary form \citep{bf93}
and there is no general consensus on extended forms. 
Combining the two equations with different timescales,
the RRMHD system inevitably becomes problematic,
especially in its impulsive phase. 
Thus, a further development of Ohm's law (e.g., \citet{ged96}) or
the reconstruction of the entire RRMHD system (e.g., \citet{koide09})
is highly desirable.

In order to study large-scale problems
which contain reconnection in arbitrary locations,
we can no longer use the localized resistivity at a fixed location.
We do need good physical or phenomenological anomalous resistivity models.
Our attempt is a first step toward this direction, and
the current-dependent resistivity model did exhibit different system evolution. 
This clearly tell us that
the modeling of the effective resistivity,
a long-standing problem of the entire reconnection research,
is very important
for RRMHD modeling of relativistic magnetic reconnection as well.

In summary,
we explored an RRMHD model of relativistic magnetic reconnection.
We corroborated earlier works and further studied fine plasmoid structures. 
Three different resistivity models are examined and they showed different system evolutions. 
It is crucially important to model an effective resistivity
for RRMHD modeling of magnetic reconnection.


\end{document}